\documentstyle[12pt]{article}
\newcommand{\be}{\begin{equation}}
\newcommand{\bea}{\begin{eqnarray}}
\newcommand{\eea}{\end{eqnarray}}
\newcommand{\ba}{\begin{array}}
\newcommand{\ea}{\end{array}}
\newcommand{\ee}{\end{equation}}

\expandafter\ifx\csname mathbbm\endcsname\relax

\else

\fi
\begin{document}

\begin{titlepage}
\begin{flushright}
IP/BBSR/2002-11\\
IPM/P-2002/009\\
hep-th/0205134
\end{flushright}

\vspace{5mm}
\begin{center}
{\Large {\bf D-Brane Solutions from New Isometries of pp-Waves}\\}
\vspace{16mm}

{Mohsen Alishahiha$^a$\footnote{alishah@theory.ipm.ac.ir} 
and Alok Kumar$^b$\footnote{kumar@iopb.res.in}}\\
\vspace{2mm}
{\em $^a$ Institute for Studies in Theoretical Physics and Mathematics (IPM)\\
P.O. Box 19395-5531, Tehran, Iran\\ }
\vspace{2mm}
{\em $^b$ Institute of Physics\\ 
Bhubaneswar 751 005, India\\}
\vspace{2mm}
\end{center}

\begin{abstract}
We use recently proposed translations isometries of pp-waves to
construct $D4$ and $D3$ brane solutions, using $T$-duality 
transformations, in exactly solvable pp-wave background originating from 
$AdS_3\times S^3$ geometry. A unique property of the new brane
solutions is the breaking of $SO(10-p-1)$ symmetry in the transverse 
direction of the branes due to the presence of constant $NS-NS$ and 
$R-R$ background fluxes. We verify that the our 
`localized' solutions satisfy the field equations and explicitly present the 
corresponding Killing spinors. We also show the connection of our results to 
certain M5-branes in pp-wave geometry.
 
\end{abstract}
\end{titlepage}

\newpage
\section{Introduction}

A good understanding of $D$-brane solutions is of great importance in 
understanding the strong coupling aspects of string theory,
as well as that of gauge theories through a string theory/gauge 
theory duality. In this connection, pp-wave backgrounds 
\cite{gueven,amati,tseytlin,seng,guev2}
have been of much interest, since they provide 
examples of duality \cite{malda}
within the context of exactly solvable 
string theories \cite{matsaev,hull,blau,russo}. This intriguing
proposal has been studied by many authors \cite{GO}-\cite{last}

In view of these developments, we investigate
the D-brane solutions\cite{DP,KNS,SK} (see also \cite{lee, Bill}
in pp-wave background further, using the 
new isometries of pp-waves proposed in \cite{MIC}, and 
obtain new D-brane solutions. 
In the case of pp-waves originating from the $AdS_3\times S^3$ 
geometry in a Penrose limit\cite{KNS}, we use the isometries of 
pp-waves proposed in \cite{MIC},
to construct $D3$ and $D4$ brane solutions, starting from the 
$D5$ solution of \cite{KNS}. A unique property of these solutions
is the breaking of $SO(10-p-1)$ symmetry along the transverse 
directions of the branes due to the presence of constant 
$NS-NS$ and $R-R$ fluxes. Due to this, the construction of the 
localized solution from the delocalized one, obtained from 
the $T$-duality transformations on the $D5$-brane solution 
is not immediately obvious. We however explicitly verify that our
localized solutions indeed satisfy the field equations.
We also show the stability of 
these solutions by explicitly finding the Killing spinors.

The plan of this letter is as following: In section-2 starting from
D5-brane solution in the PP-wave background, we will 
obtain the supergravity solution of other D-branes by making
use of T-duality. In section-3 we shall study the supersymmetric 
properties of the solutions. Section-4 is devoted to discussion and 
comments.

\section{Supergravity solution of branes in PP-wave background}

Consider a system of $N$ D5-branes solution in the PP-wave 
of type IIB string theory \cite{KNS}
\bea
ds^2&=&f^{-{1\over 2}}\left(2dx^+dx^--\mu^2 x_i^2(dx^+)^2+dx_i^2\right)
+f^{{1\over 2}}\left(dr^2+r^2d\Omega_3^2\right),\cr &&\cr
e^{2\phi}&=&f^{-1},\;\;\;\;\;\; \;\;\;\;\;\;\;\;F_{+12}=F_{+34}=2\mu, 
\cr &&\cr
F_{mnl}&=&\epsilon_{lmnp}\partial_{p}f,\;\;\;\;\;\;
f=1+{Ng_sl_s^2\over r^2},
\label{D5}
\eea
where $F$'s are RR 3-formes and $i=1,2,3,4$. This solution has
$SO(2)\times SO(2)\times SO(4)$ symmetry.

In order to find the Dp-branes solution for $p<5$, we apply T-duality on 
the D5-brane solution (\ref{D5}). To do this, it is useful to make the 
following change of coordinates:\cite{MIC}
\bea
x^+&=&{\hat x}^+,\;\;\;\;\;\;\;x^-={\hat x}^--\mu {\hat x}_1
{\hat x}_2,\;\;\;\;\;x_I={\hat x}_I,\;\;\;\;{\rm for}\;I=3,4\cr
&&\cr
x_1&=&{\hat x}_1\cos(\mu {\hat x}^+)-{\hat x}_2\sin(\mu {\hat x}^+),
\;\;\;\;
x_2={\hat x}_1\sin(\mu {\hat x}^+)+{\hat x}_2\cos(\mu {\hat x}^+).
\label{COO}
\eea
Then the D5-brane solution (\ref{D5}) takes the form:
\bea
ds^2&=&f^{-{1\over 2}}\left(2d{\hat x}^+d{\hat x}^-
-\mu^2 {\hat x}_I^2(d{\hat x}^+)^2-4\mu {\hat x}_2d{\hat x}_1d{\hat x}^+
+d{\hat x}_i^2\right)
+f^{{1\over 2}}\left(dr^2+r^2d\Omega_3^2\right),\cr &&\cr
e^{2\phi}&=&f^{-1},\;\;\;\;\;\; \;\;\;\;\;\;\;\;F_{+12}=F_{+34}=2\mu, 
\cr &&\cr
F_{mnl}&=&\epsilon_{lmnp}\partial_{p}f,\;\;\;\;\;\;
f=1+{Ng_sl_s^2\over r^2}.
\eea
In this coordinate system, ${\partial \over \partial {\hat x}_1}$ is a 
manifest isometry. This is the direction one then makes compact and 
performs T-duality along it. By making use of the T-duality rules 
\cite{TDUAL}, the T-dual background of
the above solution reads:
\bea
ds^2&=&f^{-{1\over 2}}\left(2d{\hat x}^+d{\hat x}^-
-\mu^2[{\hat x}_I^2+4{\hat x}_2^2](d{\hat x}^+)^2+d{\hat x}^2_2
+d{\hat x}_I^2\right)
+f^{{1\over 2}}\left(d{\hat x}_1^2+dr^2+r^2d\Omega_3^2\right),\cr &&\cr
e^{2\phi}&=&f^{-{1\over 2}},\;\;\;\;\;\; \;\;\;\;\;\;\;\;
F_{+2}=F_{+134}=2\mu, 
\;\;\;\;\; B_{+1}=2\mu {\hat x}_2\cr &&\cr
F_{1mnl}&=&\epsilon_{1mnlp}\partial_{p}f,\;\;\;\;\;\;
f=1+{Ng_sl_s^2\over r^2}.
\label{smearedD4}
\eea
This should corresponds to the smeared D4-brane in the type 
IIA string theory
on the PP-wave. The localized D4-brane solution can be easily 
found as following:
\bea
ds^2&=&f^{-{1\over 2}}\left(2d{\hat x}^+d{\hat x}^-
-\mu^2[{\hat x}_I^2+4{\hat x}_2^2](d{\hat x}^+)^2+d{\hat x}^2_2
+d{\hat x}_I^2\right)
+f^{{1\over 2}}\left(dr^2+r^2d\Omega_4^2\right),\cr &&\cr
f&=&1+\frac{c_4 Ng_sl_s^3}{r^3},\;\;\;\;F_{mnlp}=\epsilon_{mnlpq}
\partial_{q}f,
\label{D4}
\eea
other fields remain the same as the one in smeared solution of  
(\ref{smearedD4}), with $f$ given in (\ref{D4}). We have explicitly 
verified that the solution presented in eqn.(\ref{D4}) satisfies
the type IIA field equations. In this context, we point out that 
$\mu$ dependence appears nontrivially only in the graviton field 
equations with $(++)$ components. For example, in the expression for
the Ricci tensor $R_{++}$, one obtains a constant contribution of 
$6\mu^2$, which is canceled by equal contributions coming from 
the background $R-R$ 2-form and 4-form fluxes, as well as 
$NS-NS$ background 3-form flux of the above pp-wave solution.
Remaining nonconstant terms in $R_{++}$ are canceled by terms 
coming from the dilaton contribution, as well as the 4-form field
strength coupling to the $D4$-brane, since $f$ satisfies the 
Greens function equation in the five dimensional transverse space. 
The above solution has $SO(2)\times SO(5)$ symmetry for the metric,
which is then broken by the constant field strengths associated with 
various p-form fields. 

We can now proceed to generate a $D3$-brane
solution in a pp-wave background. One then  
again makes the
change of coordinates similar to eqn. (\ref{COO}), but in the 3
and 4 directions, {\it i.e.} 
\bea
{\hat x}^+&=&y^+,\;\;\;\;\;\;\;{\hat x}^-= y^--\mu y_3
y_4,\;\;\;\;\;{\hat x}_2=y_2,\cr
&&\cr
{\hat x}_3&=&y_3\cos(\mu y^+)-y_2\sin(\mu y^+),\;\;\;\;
{\hat x}_4=y_3\sin(\mu y^+)+y_4\cos(\mu y^+).
\eea
In this system of coordinates, the metric in (\ref{D4}) reads:
\bea
ds^2&=&f^{-{1\over 2}}\left(2dy^+dy^--4\mu^2y_2^2(dy^+)^2-
4\mu y_4dy_3dy^+
+dy_2^2+dy_I^2\right)\cr
&+& f^{{1\over 2}}\left(dr^2+r^2d\Omega_4^2\right).
\eea
In this coordinate system, ${\partial\over \partial y_3}$ is 
a manifest isometry.
One can now compactify this direction and perform T-duality. 
Doing so one finds:
\bea
ds^2&=&f^{-{1\over 2}}\left(2dy^+dy^-
-4\mu^2[y_2^2+y_4^2](dy^+)^2+dy^2_2
+dy_4^2\right)
+f^{{1\over 2}}\left(dy_3^2+dr^2+r^2d\Omega_4^2\right),\cr &&\cr
F_{+32}&=&F_{+14}=2\mu, 
\;\;\;\;\; B_{+1}=2\mu y_2,\;\;\;\;\;\; B_{+3}=2\mu y_4,\cr &&\cr
F_{3mnlp}&=&\epsilon_{3mnlpq}\partial_{q}f,\;\;\;\;\;\;
f=1+{c_4Ng_sl_s^3\over r^3}.
\eea
with a constant dilaton. This is the smeared D3-brane 
solution in the PP-wave
background. The localized solution is given by
\bea
ds^2&=&f^{-{1\over 2}}\left(2dy^+dy^-
-4\mu^2[y_2^2+y_4^2](dy^+)^2+dy^2_2
+dy_4^2\right)
+f^{{1\over 2}}\left(dr^2+r^2d\Omega_5^2\right),\cr &&\cr
F_{+32}&=&F_{+14}=2\mu, 
\;\;\;\;\; B_{+1}=2\mu y_2,\;\;\;\;\;\; B_{+3}=2\mu y_4\cr &&\cr
F_{mnlpq}&=&\epsilon_{mnlpqs}\partial_{s}f,\;\;\;\;\;\;
f=1+{c_3Ng_sl_s^4\over r^4}.
\label{d3soln.}
\eea
Once again, we have explicitly verified that the field equations are
satisfied by the solution in eqn.(\ref{d3soln.}). For example, 
in this case, the
constant $\mu$ dependent part in $R_{++}$ is equal to $8\mu^2$, which 
is canceled by equal contribution from the $NS-NS$ and $R-R$ 
field strengths. The non-constant part is canceled by the 5-form 
field strength of the $D3$-brane with $f$ satisfying the Greens
function equation in six dimensional transverse space: 
$x^1, x^3, x^5,..,x^8$. 

As another example, we can also find the M5-brane solution in the
PP-wave background of M-theory. To do this one can liftup the 
D4-brane solution (\ref{D4}) to M-theory. Using the relation
between 11 and 10 dimensional metric as
\be
ds^2_{11}=e^{{-2\phi\over 3}}ds^2_{10}+e^{{4\phi\over 3}}
(dx_{11}+A_{\mu}dx^{\mu})^2,
\ee
where $A_{\mu}$ is R-R one form, one can find the M5-brane solution 
from D4-brane solution (\ref{D4}). The M5-brane metric is given by
\bea
ds^2&=&f^{-{1\over 3}}\left(2d x^+d x^-
-\mu^2[ x_I^2+4x_2^2](d x^+)^2+d x^2_2
+dx_I^2+(dx_{11}-2\mu x_2dx^{+})^2\right)\cr &&\cr
&+&f^{{2\over 3}}\left(dr^2+r^2d\Omega_4^2\right),\cr &&\cr
&=&f^{-{1\over 3}}\left(2dx^+dx^--\mu^2 x_I^2(d x^+)^2
-4\mu x_2dx_{11}dx^++d x^2_2+dx_I^2+dx_{11}^2\right)\cr &&\cr
&+&f^{{2\over 3}}\left(dr^2+r^2d\Omega_4^2\right),
\eea
where $f=1+\frac{Nl_{p}^3}{r^3}$, with $l_p$ being the  11-dimensional 
Plank length.
Using a change of coordinate as inverse to the once in (\ref{COO}), 
one finds the
following result for M5-brane solution: 
\bea
ds^2&=&f^{-{1\over 3}}\left(2dx^+dx^--\mu^2[x_I^2+x_2^2+x_{11}^2](d x^+)^2+
d x^2_2+dx_I^2+dx_{11}^2\right)\cr &&\cr
&+&f^{{2\over 3}}\left(dr^2+r^2d\Omega_4^2\right),\cr &&\cr
C^{(4)}&=&2\mu (dx^{+}dx_ldx_3dx_4+dx^+dx_{l}dx_2dx_{11})+F_4, 
\eea
where $x_l$ is the a transverse direction to the brane, and $F_4$ comes from
the R-R 4-form in (\ref{D4}). This is the same solution as 
the one obtained in
\cite{SING}. One can also proceed to find the NS5-brane solution. To do this
we should first delocalize the M5-brane solution in a direction, say $x_l$ and
then compactify the solution in this direction. Doing so one finds
following type IIA NS5-brane solution in the PP-wave background:
\bea
ds^2&=&2dx^+dx^--\mu^2x_i^2(d x^+)^2+
dx_i^2+f\left(dr^2+r^2d\Omega_3^2\right),\cr &&\cr
e^{2\phi}&=&f,\;\;\;\;H_{+12}=H_{+34}=2\mu, \;\;\;\;
H_{mnp}=\epsilon_{mnpq}\partial_q f,
\label{ns5a}
\eea
where $i=1,2,3,4$ and $f=1+\frac{Nl_s^2}{r^2}$ and $H$'s are NS-NS 3-forms.
As expected, this $NS5$-brane solution matches with the 
$NS5$-brane of type IIB in pp-wave background\cite{KNS}. These results
have interesting implications for the supersymmetry of our solutions and 
we  discuss them in the next section. 

\section{Supersymmetric properties of the brane solution in PP-wave 
background}

We now first discuss the supersymmetry properties of the above solutions on 
general grounds and then go on to solve the Killing spinor equations. 
It was already shown in the previous section that our $D4$-brane
solution in eqn.(\ref{D4}) can be lifted to an $M5$-brane solution 
of \cite{SING}. Also, the full $M5$-brane solution preserves $3/8$
supersymmetry, with brane preserving $1/2$ of the background pp-wave
supersymmetry. We therefore expect that our $D4$-brane 
also preserves same amount of supersymmetries. Moreover, it was 
also shown in the previous section that the $M5$-brane solution can 
be reduced to an $NS5$-brane of type IIA theory. Now, since 
this solution is also identical to the $NS5$-brane of type IIB, we 
expect that $NS5$-brane also preserves the same supersymmetry in IIB
string theory. In fact, the identical pp-wave backgrounds for 
the above IIA NS-5 brane and the one in IIB theory \cite{KNS} are
related by a simple $T$-duality along one of the transverse
directions, as can be seen by setting the source term:
$f=1$ in the two solutions.

We now first analyze the Killing spinor equations of the 
$D5$-brane solution given in eqn.(\ref{D5}) and then go over 
to other brane solutions. To begin, we write 
the relevant supersymmetry transformations for dilatino and 
gravitino fields in type IIB supergravity in ten dimensions\cite{hassan}
in string frame metric:
\be
\delta \lambda_{\pm}  = {1\over 2}\left( \Gamma^{\mu} \partial_{\mu}\phi
               \mp {1\over 12}\Gamma^{\mu \nu \rho} H_{\mu \nu \rho}
                \right) \epsilon_{\pm} 
          + {1\over 2} e^{\phi} \left(\pm\Gamma^M F_M^{(1)}
           + {1\over 12}\Gamma^{\mu \nu \rho}
              F^{(3)}_{\mu \nu \rho}\right)\epsilon_{\mp},
\label{dilatino}
\ee
\bea
\delta \Psi^{\pm}_{\mu}  &=& \left[\partial_{\mu} + {1\over 4} 
       (\omega_{\mu \hat{a} \hat{b}} \mp 
{1\over 2} H_{\mu\hat{a}\hat{b}})\Gamma^{\hat{a}\hat{b}}
        \right] \epsilon_{\pm} \cr
  && \cr
    & +& {1\over 8}e^{\phi} \left[ \mp \Gamma^{\mu}F^{(1)}_{\mu}
     - {1\over 3!}\Gamma^{\mu \nu \rho} F^{(3)}_{\mu \nu \rho}
     \mp {1\over {2. 5!}}
\Gamma^{\mu \nu \rho \alpha \beta} F^{(5)}_{\mu \nu \rho \alpha \beta}
\right]\Gamma_{\mu}\epsilon_{\mp}.
\label{gravitino}
\eea
Using indices: $(+, -, i, a)$ to denote the ten dimensional
coordinates, with $i=1,..,4$ and $a=5,..,8$, we get the following
condition from the dilatino equation for the 
$D5$ solution given in eqn.(\ref{D5}) (hats denoting the 
corresponding tangent space coordinates):
\be
\Gamma^{\hat{a}} \epsilon_{\pm} + {1\over {3!}}
\epsilon_{\hat{a}\hat{b}\hat{c}\hat{d}}
\Gamma^{\hat{b}\hat{c}\hat{d}}\epsilon_{\mp} = 0,
\label{d5dilatino+}
\ee
\be
   \Gamma^{\hat{+}}\left(\Gamma^{\hat{1}\hat{2}} + 
\Gamma^{\hat{3}\hat{4}} \right)\epsilon_{\pm} = 0.
\label{additional}
\ee
Equation (\ref{d5dilatino+}) is the standard
supersymmetry condition for a $D5$-brane even in the flat space:
\be
\epsilon_{\pm} = \Gamma^{\hat{5}\hat{6}\hat{7}\hat{8}}\epsilon_{\mp},
\label{c-5678}
\ee
and reduces the supersymmetry to $1/2$ of the maximal one. 
Both the equations (\ref{additional}) and (\ref{c-5678}) are
in fact necessary, in order that dilatino variation vanishes.
These conditions are
also seen to emerge from the gravitino variations that we
analyze below. 

Using the $D5$-brane supersymmetry condition (\ref{d5dilatino+}),
the gravitino variation equations reduce to:
\be
\delta {\Psi^{\pm}}_+ \equiv \partial_+\epsilon_{\pm} - 
{\mu^2 x_{\hat{i}}\over 2}\Gamma^{\hat{+}\hat{i}}\epsilon_{\pm}
- {\mu \over 4}(\Gamma^{\hat{+}\hat{1}\hat{2}} + 
\Gamma^{\hat{+}\hat{3}\hat{4}})\Gamma^{\hat{-}}\epsilon_{\mp} = 0,
\label{grav+}
\ee
\be
\delta {\Psi^{\pm}}_- \equiv \partial_- \epsilon_{\pm} = 0, \label{grav-}
\ee
\be
\delta {\Psi^{\pm}}_ i \equiv \partial_i\epsilon_{\pm} - 
{\mu \over 4}(\Gamma^{\hat{+}\hat{1}\hat{2}} + 
\Gamma^{\hat{+}\hat{3}\hat{4}})\delta_{i \hat{i}}
\Gamma^{\hat{i}}\epsilon_{\mp} = 0,
\label{gravi}
\ee
\be
\delta {\Psi^{\pm}}_ a \equiv \partial_a\epsilon_{\pm} - 
{\mu \over 4}f^{1\over 2}(\Gamma^{\hat{+}\hat{1}\hat{2}} + 
\Gamma^{\hat{+}\hat{3}\hat{4}})\delta_{a \hat{a}}
\Gamma^{\hat{a}}\epsilon_{\mp} = 0.
\label{grava}
\ee
We now use eqn.(\ref{additional}) to simplify eqns. 
(\ref{grav+}) - (\ref{grava}) further. First, assuming 
$\Gamma^{\hat{+}}\epsilon_{\pm} \neq 0$ gives:
\be
   \left( 1 + \Gamma^{\hat{+}\hat{-}\hat{1}\hat{2}\hat{3}\hat{4}} \right)
   \epsilon_{\pm} = 0.
\label{pp-cond}
\ee
In writing this equation, as well as the ones give below, we have
made use of the identities: $\Gamma^{\hat{+}}\Gamma^{\hat{+}\hat{-}}
= - \Gamma^{\hat{+}}$, $\Gamma^{\hat{+}\hat{-}}\Gamma^{\hat{-}}
= - \Gamma^{\hat{-}}$ etc.. 
Now, by defining 32-component real spinors:
\bea
    \eta \equiv \pmatrix{\epsilon_+\cr \epsilon_-},
\eea
and using eqn.(\ref{pp-cond}), eight spinor equations in 
(\ref{grav+})-(\ref{grava}), following from 
$\delta {\Psi^{\pm}}_{\mu}=0$ can be written as:
\be
\partial_+\eta - 
{\mu^2 x_{\hat{i}}\over 2}\Gamma^{\hat{+}\hat{i}}\eta 
- {\mu \over 2}\Gamma^{\hat{+}\hat{1}\hat{2}} \Gamma^{\hat{-}}\otimes
\sigma_1 \eta = 0,
\label{grav+1}
\ee
\be
\partial_- \eta = 0, \label{grav-1}
\ee
\be
\partial_i\eta 
- {\mu \over 2}(\Gamma^{\hat{+}\hat{1}\hat{2}} )\delta_{i \hat{i}}
\Gamma^{\hat{i}}\otimes \sigma_1 \eta = 0,
\label{gravi1}
\ee
\be
\partial_a \eta  = 0,
\label{grava1}
\ee
where $\sigma_1$ is the Pauli matrix, mixing the two components
$\epsilon_+, \epsilon_-$ of $\eta$. Moreover, using 
eqns. (\ref{d5dilatino+}), (\ref{c-5678}), (\ref{pp-cond}) 
and the fact that both $\epsilon_{\pm}$ are spinors of same 
space-time helicity, one has: $\eta = \sigma_1 \eta$, or 
both components of $\eta$ are to be equal. However, one should 
take into account that such a condition is not independent 
with respect to what has been written earlier in 
eqns.(\ref{d5dilatino+}) and (\ref{pp-cond}). 

The solution of the above Killing spinor equation is then found 
in a similar way as in \cite{hull},\cite{meessen} and is given 
as:
\be
\eta = \left(1 + {\mu\over 2}
\Gamma^{\hat{+}\hat{1}\hat{2}}x^i \delta_{i \hat{i}}\Gamma^{\hat{i}}    
\right)\xi (u),
\label{k-spinor}
\ee
\be
\xi (u) = exp.\left({\mu\over 2}\Gamma^{\hat{+}\hat{1}\hat{2}}
\Gamma^{\hat{-}} u\right)\xi_0,
\label{def-xi}
\ee
with $\xi_0$ being a constant spinor.
The Killing spinors found in 
eqns. (\ref{k-spinor}), (\ref{def-xi}) are constrained by the 
conditions (\ref{pp-cond}) and (\ref{d5dilatino+}). The fact that 
the solutions for both these equations, (\ref{c-5678})
(\ref{pp-cond}), are also given by 
(\ref{k-spinor}) and (\ref{def-xi}), is due to the fact
that matrices $\Gamma^{\hat{5}\hat{6}\hat{7}\hat{8}}$, as well
as $\Gamma^{\hat{+}\hat{-}\hat{1}\hat{2}\hat{3}\hat{4}}$, both 
commute with the product of Gamma matrices appearing in 
eqns.(\ref{k-spinor}) and (\ref{def-xi}).
Before doing the final count of the number of supersymmetries that 
are preserved by our background fields, we now examine the possibility of
having more solutions of the Killing spinor equations. Later on, 
we also discuss the connection of our Killing spinors with the 
`normal' and `supernumerary' Killing spinors in 
\cite{pope,SING}.

Additional solutions of both the dilatino and the
gravitino variations are possible by using projections:
\be
\Gamma^{\hat{+}}\epsilon_{\pm} = 0.
\label{condg+}
\ee 
The condition (\ref{additional})
is now trivially satisfied and eqns.(\ref{grav+})-(\ref{grava}) are
replaced by:
\be
\partial_+\epsilon_{\pm} 
- {\mu \over 2}(\Gamma^{\hat{1}\hat{2}} + 
\Gamma^{\hat{3}\hat{4}})\epsilon_{\mp} = 0,
\label{grav+2}
\ee
\be
\partial_- \epsilon_{\pm} = 0, 
\>\>\>\>
\partial_i\epsilon_{\pm} = 0,
\>\>\>\>
\partial_a\epsilon_{\pm} = 0.
\label{grava2}
\ee
However, before solving these equations, we notice by using 
\be
\Gamma^{\hat{+}} = {1\over 2} \Gamma^{\hat{+}}(1 - 
\Gamma^{\hat{+}\hat{-}}),
\label{defgama+}
\ee
that a part of the 
solutions of eqn. (\ref{condg+}) already coincide with the solutions
of the type in eqns. (\ref{k-spinor}), (\ref{def-xi}): further 
constrained by the projection condition:
(\ref{pp-cond}). As a result, only those 
spinors in (\ref{pp-cond}) are relevant which 
satisfy:
\be
\Gamma^{\hat{1}\hat{2}\hat{3}\hat{4}}
\epsilon_{\pm} = \epsilon_{\pm}.
\label{supernum}
\ee
In any case, the other projection:
\be
\Gamma^{\hat{1}\hat{2}\hat{3}\hat{4}}
\epsilon_{\pm} = - \epsilon_{\pm},
\label{supernum2}
\ee
is inconsistent with eqn. (\ref{pp-cond}), as one will then have
$\Gamma^{\hat{+}}\epsilon_{\pm} = 0$, using eqn. (\ref{defgama+}). 
Total number 
of supersymmetries preserved by the background 
pp-wave will be $3/4$ of 
the maximal ones, since eqn. (\ref{grav+2}) can be solved 
easily after imposing either (\ref{supernum}) or (\ref{supernum2}). 
In addition, the brane breaks another $1/2$ supersymmetry given by 
eqns.(\ref{d5dilatino+}), (\ref{c-5678}), so that total solution 
(\ref{D5}) preserves $3/8$ of the maximal supersymmetry. 

We now comment on the relation of the Killing spinor solutions 
discussed above, 
with the ones in \cite{hull}, \cite{SING}. It is now apparent that
the solutions of the Killing equations with projection
(\ref{condg+}) are the `normal' Killing spinors
of \cite{SING}, whereas the ones given by solution of the 
Killing spinor equations, with condition (\ref{pp-cond})
are the `supernumerary' Killing spinors, since independent 
solutions of this projection condition is also a solution of
another projection condition given by (\ref{supernum}).

To summarize, combining results presented in eqns.(\ref{d5dilatino+}), 
(\ref{c-5678}),
(\ref{pp-cond}), (\ref{k-spinor}), (\ref{def-xi}) and 
(\ref{condg+}), (\ref{supernum}), we conclude that 
the $D5$-brane solution in 
eqn.(\ref{D5}) preserves $1/2$ of the `background' supersymmetry. 
The pp-wave background itself preserves $3/4$ of the maximal 
supersymmetries. We have also examined the pp-wave `background' 
geometry of the IIB $NS5$-brane. The analogy with the Killing 
equations of \cite{SING} holds there as well and once again gives
$3/4$ supersymmetry.

The supersymmetry analysis for the $D3$-brane solution presented in 
eqn.(\ref{d3soln.}) of this paper is very similar to the one 
given above. The dilatino variation now gives:
\be
\delta \lambda_{\pm} \equiv \mp {\mu\over 2} f^{1\over 4}
\left(\Gamma^{\hat{+}\hat{1}\hat{2}} 
               + \Gamma^{\hat{+}\hat{3}\hat{4}}\right)\epsilon_{\pm}
 + {\mu\over 2} f^{1\over 4}
\left(\Gamma^{\hat{+}\hat{3}\hat{2}} 
               + \Gamma^{\hat{+}\hat{1}\hat{4}}\right)\epsilon_{\mp}
          = 0,
\label{d3dilatino}               
\ee
and is solved by imposing condition:
\be
\Gamma^{\hat{+}}\epsilon_{\pm} = 0.
\label{gamma+}
\ee
The need to impose condition (\ref{gamma+}) 
for the $D3$-brane solution becomes 
apparent while writing the gravitino variation equations for 
components $(i=2, 4)$ below, 
as the $NS-NS$ 3-form flux contributes 
a term which can be consistently
set to zero by imposing the above 
condition. More explicitly, we have (using eqn.(\ref{gamma+})):
\bea
\partial_+ \epsilon_{\pm} - {1\over 8} {{f,}_{\hat{a}}\over f^{3\over 2}}
\Gamma^{\hat{-}\hat{a}}\epsilon_{\pm} 
\mp {\mu\over 2} (\Gamma^{\hat{1}\hat{2}} 
+\Gamma^{\hat{3}\hat{4}})\epsilon_{\pm}
- {\mu \over 4}(\Gamma^{\hat{+}\hat{3}\hat{2}} + 
\Gamma^{\hat{+}\hat{1}\hat{4}})\Gamma^{\hat{-}}\epsilon_{\mp} \cr
\cr
\mp {1\over 8} {1\over 5!}
\Gamma^{\hat{a_1}...\hat{a_5}}\epsilon_{\hat{a}_1...\hat{a}_5 a}
{f_{,a}\over f^{3\over 2}}\Gamma^{\hat{-}}\epsilon_{\mp} = 0,
\label{grd3+}
\eea
\be
\partial_- \epsilon_{\pm} = 0,\>\>\>\>
\partial_i \epsilon_{\pm} = 0,\>\>\>(i=2, 4),
\label{grd3i}
\ee
\be
\partial_a \epsilon_{\pm} + {1\over 8} {f_{,\hat{c}}\over f}
\delta_{a \hat{b}}\Gamma^{\hat{b}\hat{c}}\epsilon_+
\mp {1\over 8}{1\over 5!}
\Gamma^{\hat{a_1}...\hat{a_5}}\epsilon_{\hat{a}_1...\hat{a}_5 b}
{f_{,b}\over f} \delta_{a \hat{a}}\Gamma^{\hat{a}}\epsilon_{\mp} =0, 
\>\>\> (a = 1, 3, 5,..,8).
\label{grd3a}
\ee
Then the $D3$-brane supersymmetry condition: 
\be
 \epsilon_+ = \Gamma^{\hat{1}\hat{3}\hat{5}...\hat{8}}\epsilon_-
\label{d3cond}
\ee
implies that all the Killing spinor equations are satisfied, provided
one also imposes
\be
(\Gamma^{\hat{3}\hat{2}} + \Gamma^{\hat{1}\hat{4}})\epsilon_{\pm} = 0.
\label{extra}
\ee
or
\be
(\Gamma^{\hat{1}\hat{2}} + \Gamma^{\hat{3}\hat{4}})\epsilon_{\pm} = 0.
\label{extra2}
\ee
Equations (\ref{gamma+}),(\ref{extra}) (or (\ref{extra2})) 
give the Killing spinor
conditions on the background pp-wave and the condition in 
eqn.(\ref{d3cond}) is the supersymmetry breaking due to the 
presence of $D3$-brane. 
Imposing all these conditions, one finally observes 
using (\ref{extra}), that 
$\epsilon_{\pm}$ are given as:
\be
\epsilon_{\pm} = exp ({\mu\over 2}\Gamma^{\hat{1}\hat{2}} u)
                 \epsilon_{\pm}^0.
\label{d3killing}
\ee
This is a valid solution, as the projection matrices 
$\Gamma^{\hat{+}}$ as well as $\Gamma^{\hat{1}\hat{2}\hat{3}\hat{4}}$, 
both commute with the term in the exponential in
eqn.(\ref{d3killing}). Combining these with another set of spinors
coming from (\ref{extra2}),
we have $1/2$ supersymmetry for the pp-wave
background metric in eqn.(\ref{d3soln.}). In addition, the brane once again 
breaks another $1/2$ supersymmetry, leading to $1/4$ supersymmetry for 
the full solution. As an observation, we mention that 
additional Killing spinors are possible, when one
of $NS-NS$ background fields has an opposite signature than the ones
in (\ref{d3soln.}), as the dilatino equation (\ref{d3dilatino})
can be satisfied by imposing a projection different from the one
in (\ref{gamma+}):
$\Gamma^{\hat{3}\hat{2}\hat{2}\hat{4}}\epsilon_{\pm} =
\epsilon_{\pm}$. This condition also simplifies the gravitino 
equations, leading now to $3/4$ supersymmetry for the background.
Finally, 
the above exercise can be repeated for the $D4$-brane
solution in eqn.(\ref{D4}) as well and we once again expect the 
solution to be stable.

\section{Discussions}

In this paper we have found the supergravity solution of different 
branes in the PP-wave background using toroidal isometries of pp-waves, 
following Michelson\cite{MIC}. To find these solutions, we have started
from the known D5-brane solution and then used S- and T-duality. We have
also studied the supersymmetric properties of these branes where we showed 
that they preserve half of the background supersymmetry.

Having supergravity solution of different branes one could proceed to study
the worldvolume theory of them. For D7-branes this has been studied
in \cite{DP} where it has been shown that the CS-term plays an important
role in order to get the correct mass for gauge field. To study other
branes we note, however, that it is also crucial to consider the Myers
term\cite{myers} as well. In particular, for D3-brane case it seems
that it is the Myers term which provides the correct mass for the
adjoint scalars. We however leave this as a future exercise.

Another interesting feature of these supergravity solutions would be
to extend the original Maldacena's conjecture for the string
theory to the one in  the PP-wave background. Namely one might 
expect that the theory on the worldvolume of D3-brane in the 
PP-wave background decouples from the
bulk gravity, providing a dual description of the string theory on the 
near-horizon limit of (\ref{d3soln.}). 

{\bf Note added:} While we were preparing our paper for submission
we received the paper \cite{Bain} where the supergravity solution of
branes have also been studied.

\vskip .5cm

{\bf Acknowledgments}

We would like to thank S. Parvizi for useful discussions.

\end{document}